\documentclass[aps,groupedaddress,amsmath,amssymb,longbibliography]{revtex4-1}
\usepackage{graphicx}
\usepackage{color}
\usepackage{bm}
\usepackage{hyperref}
\usepackage{soul}
\usepackage{lineno}

\DeclareMathOperator{\tr}{Tr}

\begin{document}
\title{Tunable structure and dynamics of active liquid crystals}

\author{Nitin Kumar$^{1,2}$}
\thanks{Equal contribution}
\author{Rui Zhang$^{3}$}
\thanks{Equal contribution}
\author{Juan J. de Pablo$^{3,4}$}
\email{depablo@uchicago.edu}
\author{Margaret L. Gardel$^{1,2,5}$}
\email{gardel@uchicago.edu}
\affiliation{
	$^1$James Franck Institute, The University of Chicago, Chicago, Illinois 60637, USA \\
	$^2$Department of Physics, The University of Chicago, Chicago, Illinois 60637, USA \\
	$^3$Institute for Molecular Engineering, The University of Chicago, Chicago, Illinois 60637, USA \\
	$^4$Institute for Molecular Engineering, Argonne National Laboratory, Lemont, Illinois 60439, USA\\
	$^5$Institute for Biophysical Dynamics, The University of Chicago, Chicago, Illinois 60637, USA
	}
\date{\today}
\begin{abstract}
Active materials are capable of converting free energy into directional motion, giving rise to striking dynamical phenomena. Developing a general understanding of their structure in relation to the underlying non-equilibrium physics would provide a route towards control of their dynamic behavior and pave the way for potential applications. The active system considered here consists of a quasi-two-dimensional sheet of short ($\approx$ 1 $\mu$m) actin filaments driven by myosin-II motors. By adopting a concerted theoretical and experimental strategy, new insights are gained into the non-equilibrium properties of active nematics over a wide range of internal activity levels. In particular, it is shown that topological defect interactions can be led to transition from attractive to repulsive as a function of initial defect separation and relative orientation. Furthermore, by examining the +1/2 defect morphology as a function of activity, it is found that the apparent elastic properties of the system (the ratio of bend-to-splay elastic moduli) are altered considerably by increased activity, leading to an effectively lower bend elasticity. At high levels of activity, the topological defects that decorate the material exhibit a liquid-like structure, and adopt preferred orientations depending on their topological charge. Taken together, these results suggest that it should be possible to tune internal stresses in active nematic systems with the goal of designing out-of-equilibrium structures with engineered dynamic responses.

\end{abstract}
\maketitle

\section*{Introduction}
Materials that contain mechanochemically active constituents are broadly referred to as active matter and are ubiquitous in natural \cite{herd,fish}, biological \cite{needleman_dogic_nrm2017} and physical \cite{granular1,paxton_jacs2004, Alicea2005}  systems. The internal stresses that activity generates result in materials that can spontaneously flow and deform over macroscopic length scales \cite{RMP}. A fundamental question in active matter physics is how internal energy impacts the structure, mechanics and dynamics of a material that is out of thermodynamic equilibrium.

Structured fluids are a particularly rich system in which to explore these questions.
On the one hand, nematic liquid crystals (LCs) can be used to manipulate active matter \cite{aranson_pnas2014,genkin_prx2017,peng_science2016}. On the other hand, activity may destroy orientational order of LCs and lead to generation of defect-pairs and spontaneous flows\cite{simha}.
This behavior has been experimentally realized in vibrated granular matter \cite{vjScience}, dense microtubule solutions driven by kinesin motors \cite{dogic_nature2012}, bacterial suspensions\cite{bacteria}, and cell colonies\cite{saw_nature2017,kawaguchi_nature2017}.
In the microtubule-kinesin active nematics of relevance to this work, recent efforts have sought to alter defect structure using confinement\cite{keber_science2014} or surface fields\cite{guillamat_pnas2016}, and have sought to characterize transport properties such as viscosity \cite{guillamat_pre2016}, elasticity and active stresses \cite{ellis_natphys2018}.
Despite this increased interest, foundational questions regarding the role of activity on characteristic length scales of active flows, or the nature of defect-pair interactions far from equilibrium, remain unanswered. Addressing these questions might enable design and engineering of new classes of active and adaptive materials.

Here, we introduce a nematic LC composed of short actin filaments driven into an active state by myosin-II motors. We first demonstrate that the long-time clustering dynamics of myosin motors can be exploited to probe the LC over a range of active stresses.  We measure changes in the LC's orientational and velocity correlation lengths as a function of motor density and find that these are consistent with theoretical calculations of nematic LCs with varying levels of internal stress. We then use the morphology of +1/2 defects to show, in both experiments and simulations, that increased activity reduces the LC's bend elasticity relative to its splay elasticity. Thus, the degree to which an LC with known mechanics is driven out of equilibrium can be ascertained by the +1/2 defect morphology. We further demonstrate that varying internal activity can completely alter defect interaction, turning it from attractive to effectively repulsive. The activity at which this transition occurs is found to be a function of a defect pair's initial separation and its relative orientation. To accurately capture these dynamics in simulations, contributions of both bend and splay elasticity in LC mechanics must be accounted for. We also analyze pair positional and orientational correlations of defects. Our calculations, which are confirmed by experimental observations and measurements, demonstrate that defects in active nematics exhibit liquid-like behavior. Finally, we show that there exists two preferable configurations for $\pm 1/2$ defect pairs, in contrast to like-charge defect pairs, which have single preferable configuration. These results demonstrate how internal stresses can be used to systematically change the mechanics and dynamics of LCs to enable structured liquids with tunable transport properties.

\begin{figure*}
	\centerline{\includegraphics[width=0.85\textwidth]{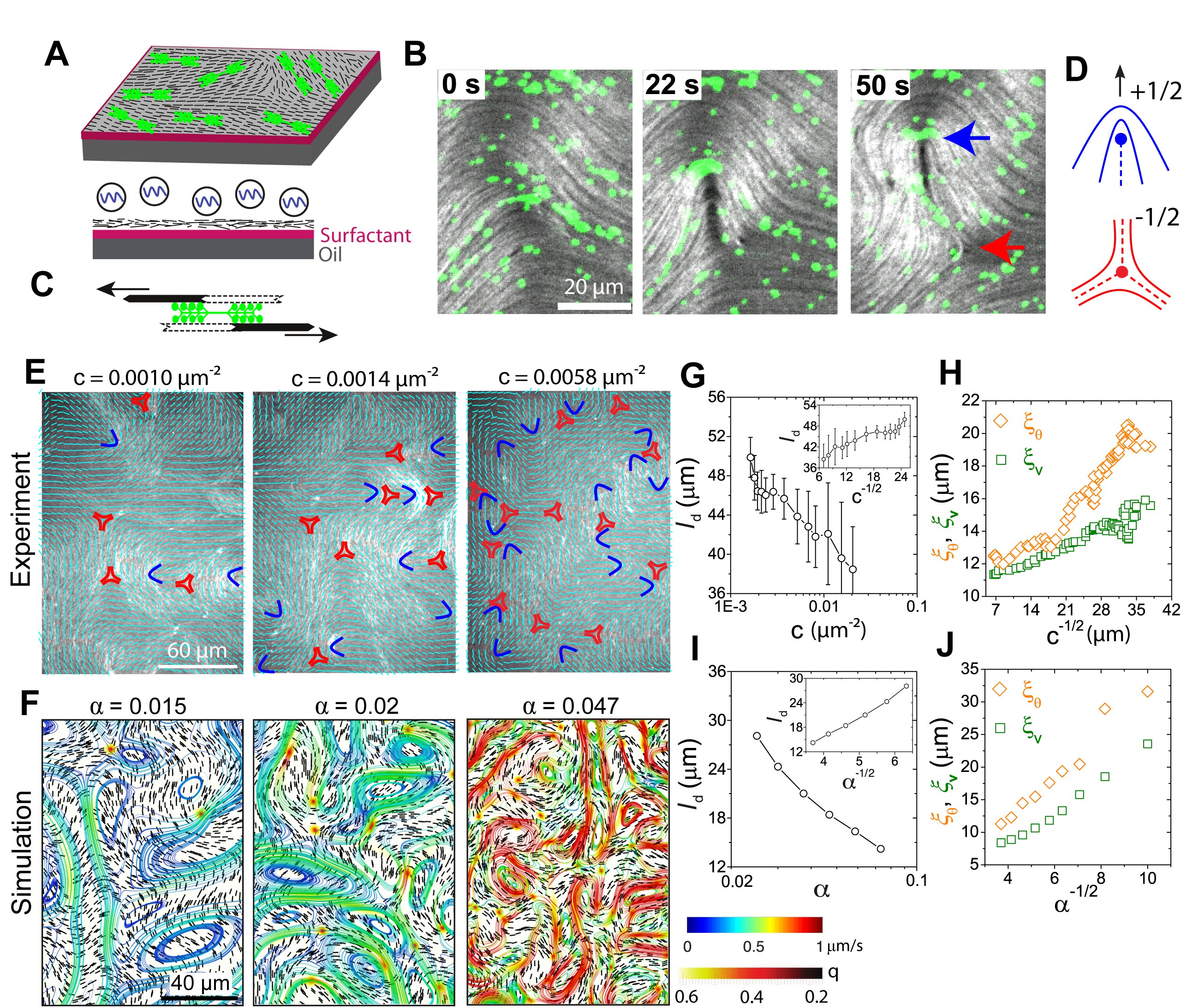}}
	\caption{ \textbf{F-actin-based active nematic LC driven by myosin-II motors}. (A) A schematic of the experimental setup. Short actin filaments (black lines) are crowded to an oil-water interface supported by a layer of surfactant molecules (magenta) to form a 2D nematic LC. After formation of the passive LC, myosin motors (green) are added. (B) Time sequence of fluorescence images of actin filaments (gray scale) and myosin motors (green) showing the generation of a $\pm$1/2 defect pair (blue, red arrows). Motor concentration, c =  0.019 $\mu m^{-2}$. Filament length, $l$ = 1 $\mu$m. (C) Schematic of two actin filaments with antiparallel polarities sliding relative each other due to the myosin-II motor activity. (D) Schematic of $\pm$1/2 defects is shown. (E) Fluorescence images of actin LC ($l$ = 1 $\mu$m) at different motor densities $c$; The director field (cyan lines) and $\pm$1/2 defects (blue and red, respectively) are overlaid. (F) Simulation snapshots of LC at different activity levels $\alpha$. Short black lines depict the local director field and curves show streamlines (color indicates speed), with warmer colors indicating higher speeds. (G) Mean defect spacing $l_d$ as a function of $c$ Inset: $l_d$ plotted against $c^{-1/2}$ (H) Orientational and velocity-velocity correlation lengths ($\xi_{\theta}$ and $\xi_v$ respectively) plotted against $c^{-1/2}$ (I) $l_d$ as a function of $\alpha$ Inset: $l_d$ plotted against $\alpha^{-1/2}$ (J) $\xi_{\theta}$ and $\xi_v$ plotted against $\alpha^{-1/2}$.}
	\label{fig1}
\end{figure*}

\section*{Results}
\subsection*{Actin-based active nematics with varying activity}

We construct an active LC formed by the semi-flexible biopolymer, F-actin, and the molecular motor, myosin II. A dilute suspension of monomeric actin (2 $\mu$M) is polymerized in the presence of a capping protein (CP) \cite{CP} to construct filaments with a mean length of $l$ $\approx$ 1 $\mu$m.  Filaments are crowded onto an oil-water interface by adding methylcellulose as a depletion agent (Fig. \ref{fig1}A), resulting in a dense film of filaments that form a two-dimensional nematic LC with an abundance of $\pm$1/2 topological defects  (Video 1) \cite{RZhang}. Myosin II assembles into bipolar filaments of several hundred motor heads that appear as near-diffraction-limited puncta in fluorescence microscopy (Fig. \ref{fig1}B, Video 2). Myosin-II filaments generate stress on anti-parallel actin filament pairs (Fig. \ref{fig1}C) to drive changes in the LC structure and dynamics, including the formation, transport, and annihilation of defect pairs (Videos 2$-$4). Creation of new $\pm$1/2 defects occurs over tens of seconds, with defects moving apart at rates of $\sim$ 0.8 $\mu$m/s (Fig. \ref{fig1}B). The direction of +1/2 defect motion indicates that the actomyosin generates extensile stresses \cite{Giomi}, consistent with previous active nematics formed with microtubule-kinesin mixtures \cite{dogic_nature2012}. This leads to the manifestation of active nematics, which, to the best of our knowledge, has not been reported in actin-based systems.

Over the course of 50 minutes, motors cluster into larger aggregates, resulting in a decrease in the myosin puncta number density (Video 2, SI Fig. \ref{s1}). Motor clustering occurs concomitantly with a gradual decrease of the instantaneous velocity of the nematic (SI Fig. \ref{s2}), suggesting decreased active stress. Interestingly, we observe only a small, local distortion of the nematic field around large motor clusters. These clusters do not localize into the defects (SI Fig. \ref{s3}), indicating that motor clustering does not impact LC structure. To explore the LC structure as the myosin puncta density, $c$, changes, we extract the nematic director field  \cite{patrick} and identify the $\pm$1/2 defects over time as the puncta density decreases from 0.02 $\mu$m$^{-2}$ to 0.0016 $\mu$m$^{-2}$ (Fig. \ref{fig1}E). The fast relaxation time of the underlying actin LC relative to the rate of change of the motor density allows one to consider the system to be in a quasi-steady state (See analysis section in methods and SI Fig. \ref{s4}).

\subsection*{Effect of activity on correlation lengths: Myosin concentration acts as an activity parameter}

To understand how internal stress drives nematic activity, we turn to a hydrodynamic model of active nematics \cite{zhang_natcomm2016}. In the model, a phenomenological free energy is written in terms of a second-order, symmetric, and traceless ${\bf Q}$-tensor: ${\bf Q}=q({\bf nn}-{\bf I}/3)$ under uniaxial condition, with $q$ the nematic scalar order parameter, $\textbf{n}$ the director field, and $\textbf{I}$ the identity tensor (see Supplementary Text). The active stress ${\bf \Pi}^a$ caused by the presence of motors is written as
$$
{\bf \Pi}^{a} = -\alpha {\bf Q},
$$
where the activity parameter $\alpha$ has units of $N/m^2$ and is related to the magnitude of the force dipole that gives rise to local active, extensile stress. Physically, the competition of active stress and elastic stress leads to the generation of new defects, a feature that is characteristic of active nematics. Indeed, if we introduce an elastic constant $K$ for the nematic material, $\alpha/K$ bears the same units as $c$ in the experiment. This is consistent with our intuition that motor number density is related to the activity of the system. One can therefore construct a natural length scale $\sqrt{K/\alpha}$ and, as discussed later, it should dictate the characteristic lengths that arises in our active nematics.

In Fig. \ref{fig1}F, we illustrate the hydrodynamic flows obtained from simulations with different levels of activity $\alpha$. As $\alpha$ is increased, the average speed increases, indicated by the warmer color of the stream lines. The $+1/2$ defects are always associated with high-velocity regions, whereas $-1/2$ defects are stagnation points. We also observe the formation of eddies, induced by the motion of defect pairs, and find that the average eddy size decreases with increased activity. The simulations also show that the distance between defect pairs in active nematics is susceptible to large fluctuations, caused by the competition between elastic forces and active stresses. The mean defect spacing decreases as a function of active stress (Fig. \ref{fig1}I) and agrees qualitatively with that observed experimentally (Fig. \ref{fig1}G). Importantly, we observe that $l_d\propto 1/\sqrt{c}\propto \sqrt{K/\alpha}$ ((Fig. \ref{fig1}G+I, insets), consistent with theoretical expectations \cite{marchettiSM}.

\begin{figure*}
	\centerline{\includegraphics[width=0.7\textwidth]{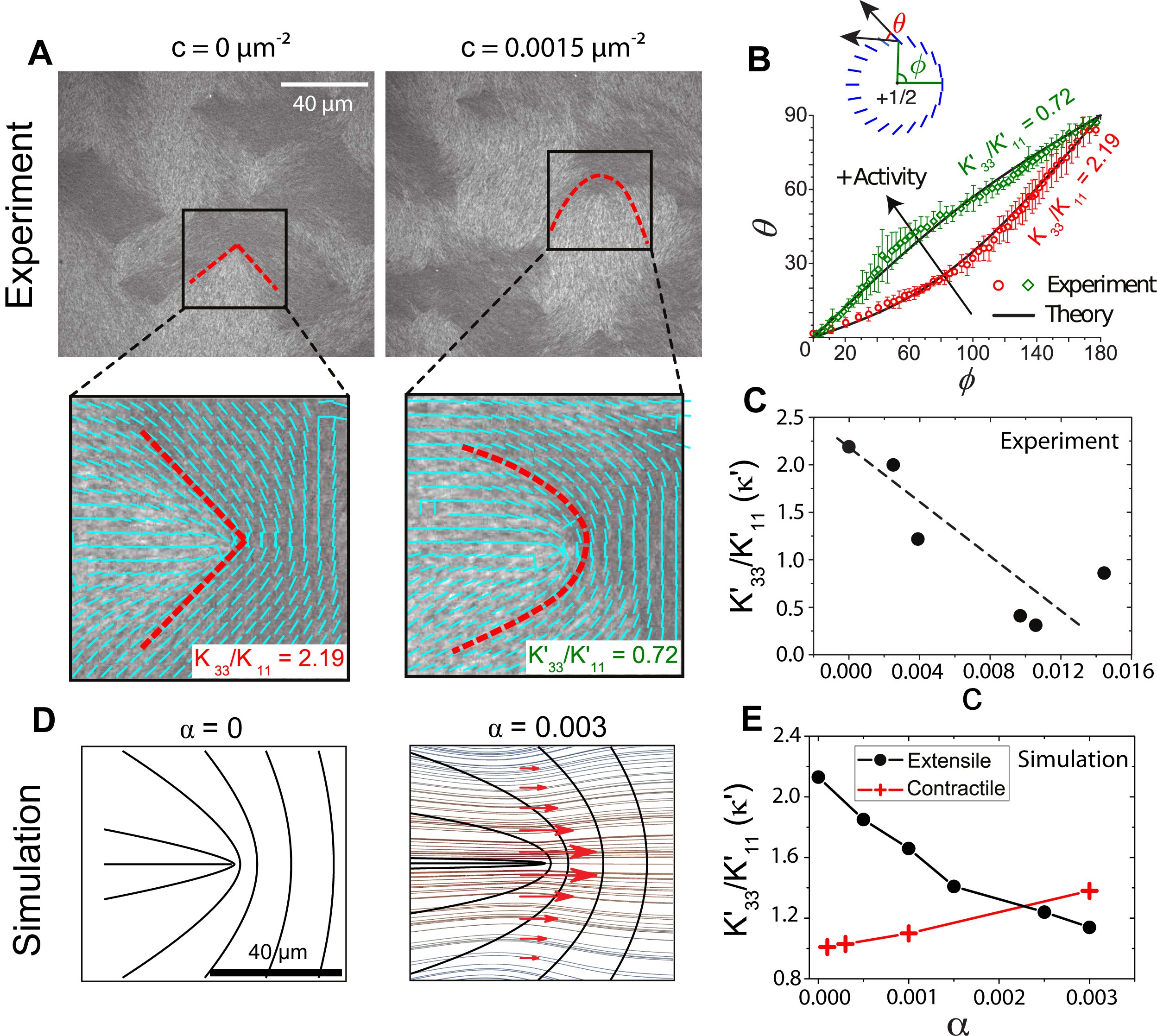}}
	\caption{\textbf{Effect of activity on defect structure and effective elasticity}. (A) Fluorescence images filaments of a passive ($c$ = 0 $\mu m^{-2}$) and active ($c$ = 0.0015 $\mu m^{-2}$) actin LC ($l$ = 2 $\mu$m). The region enclosed by the box is enlarged below and the director field (cyan lines) and defect morphology (red dashed line) are indicated. The ratio of bend ($K_{33}$) to splay ($K_{11}$) elasticity calculated from defect morphology are indicated in the bottom right. (B) Plot of $\theta$ vs. $\phi$ corresponding to experimental images of (A) for the passive (red circles) and active (green diamonds) LC. (C) Apparent elasticity $\kappa'=K'_{33}/K'_{11}$ as a function of $c$ for experimental data. Dashed line highlights the linear scaling. (D) Director field from the simulation for both passive ($\alpha$ = 0) and active ($\alpha$ = 0.003) LC.  Red arrows around the defect represent the shear flow caused by the velocity field shown in the background. (E) Apparent elasticity $\kappa'$ as a function of $\alpha$ obtained from simulations for extensile (black circles) and contractile (red crosses) stresses.}
	\label{fig2}
\end{figure*}

To further characterize the LC structure and dynamics, we measure the orientational and velocity-velocity correlation lengths, $\xi_{\theta}$ and $\xi_{v}$, respectively (See analysis section in Methods). These correlation lengths, along with $l_{d}$, have been shown to scale with activity $\xi_{\theta} \propto \xi_{v} \propto \alpha^{-1/2}$ \cite{marchettiSM}, consistent with our simulation results (Fig. \ref{fig1}J). When these correlation lengths are extracted from the experimental data, we observe that they are proportional to $c^{-1/2}$ (Fig. \ref{fig1}H). These findings serve to establish that the number density of myosin puncta is a good measure of internal active stress in the actin-based nematic. Further, the time-dependent motor clustering provides a tool to directly observe the effects of varying activity on nematic structure and dynamics.

\subsection*{Change of +1/2 defect morphology indicates lowering of effective bend-to-splay modulus ratio}

Next, we explore how changes in active stresses impact the relative balance of bend and splay energies, which is manifested in the morphology of +1/2 defects \cite{RZhang}.
Fig. \ref{fig2}A shows a fluorescence image of a passive ($c$ = 0 $\mu$m$^{-2}$) LC with average filament length, $l$ = 2 $\mu$m. For clarity, the region around a +1/2 defect has been enlarged, and the corresponding director field is shown. In a 2D nematic system, the only relevant elastic modes are splay ($K_{11}$) and bend ($K_{33}$) and their ratio, $\kappa= K_{33}/K_{11}$, dictates the morphology of $+1/2$ defects \cite{hudson1989frank,zhou_natcomm2017}. Qualitatively, the ``V-shaped'' defect morphology can be understood by the relative dominance of the bend elasticity ($K_{33}$) to the splay elasticity ($K_{11}$). We quantify the defect morphology by circumnavigating the defect and plotting the angle the director field subtends with the tangent, $\theta$, as a function of the angular coordinate $\phi$ (Fig. \ref{fig2}A+B) averaged over a radial distance from the core where it remains relatively constant \cite{RZhang}. These results are then fitted with a theoretical expression to extract a value of $\kappa$= 2.19 \cite{RZhang}.

In the presence of activity ($c$ = 0.0015 $\mu$m$^{-2}$), the defect morphology changes from ``V-shaped'' to ``U-shaped'' (Fig. \ref{fig2}A and Video 5). This is reflected in the $\theta(\phi)$ plot, from which we calculate $\kappa'$ = 0.72 (Fig. \ref{fig2}B). By analyzing defects over a wide range of $c$, we find that $\kappa$ decreases linearly with $c$ (Fig. \ref{fig2}C). Thus, increased activity results in a lower value of effective bend modulus relative to the splay modulus, consistent with the fact that extensile active nematics are unstable to bend distortion \cite{simha_prl2004,Giomi}. In other words, activity reduces the elastic penalty of nematic bend modes. The net effect is then a reduction of the effective bend modulus of the LC as activity increases.

The change of defect morphology induced by activity can also be explained in terms of hydrodynamic effects. We show in Fig.~\ref{fig2}D the flow pattern obtained from our simulations that are associated with the motion of a $+1/2$ defect as $\alpha$ is increased from 0 (left) to 0.003 (right). There are shear flows on the two sides of the symmetry axis of the defect, with which the director field of a nematic LC tends to align, assuming a flow aligning nematic \cite{deGennes_book}. With such flow-aligning effect, the surrounding director field becomes more horizontal. This leads to the defect morphology becoming more ``U-shaped'' and a lower apparent bend modulus as the internal stress increases (Fig. \ref{fig2}E, circles). Further analysis, detailed in SI Fig. \ref{s5}, shows that even though this effect is amplified by hydrodynamic flow-alignment\cite{lavrentovich_book}, activity-promoted bending is present even when the coupling to flow is turned off. In the simulations, we can change the sign of the active stress from extensile to contractile.  These calculations predict that contractile stresses lower the effective splay elasticity, resulting in a tendency for the defect to become more ``V-shaped'' (Fig. \ref{fig2}E, crosses). Importantly, these observations help establish that the defect morphology provides a direct reflection of the extent to which the LC is driven out of equilibrium. The nature of the defects' morphological change also provides a simple visual marker to differentiate between the extensile or contractile nature of active stresses.

\subsection*{Activity as a means to switch the interaction between $\pm$1/2 defects}

\begin{figure*}
	\centerline{\includegraphics[width=0.8\textwidth]{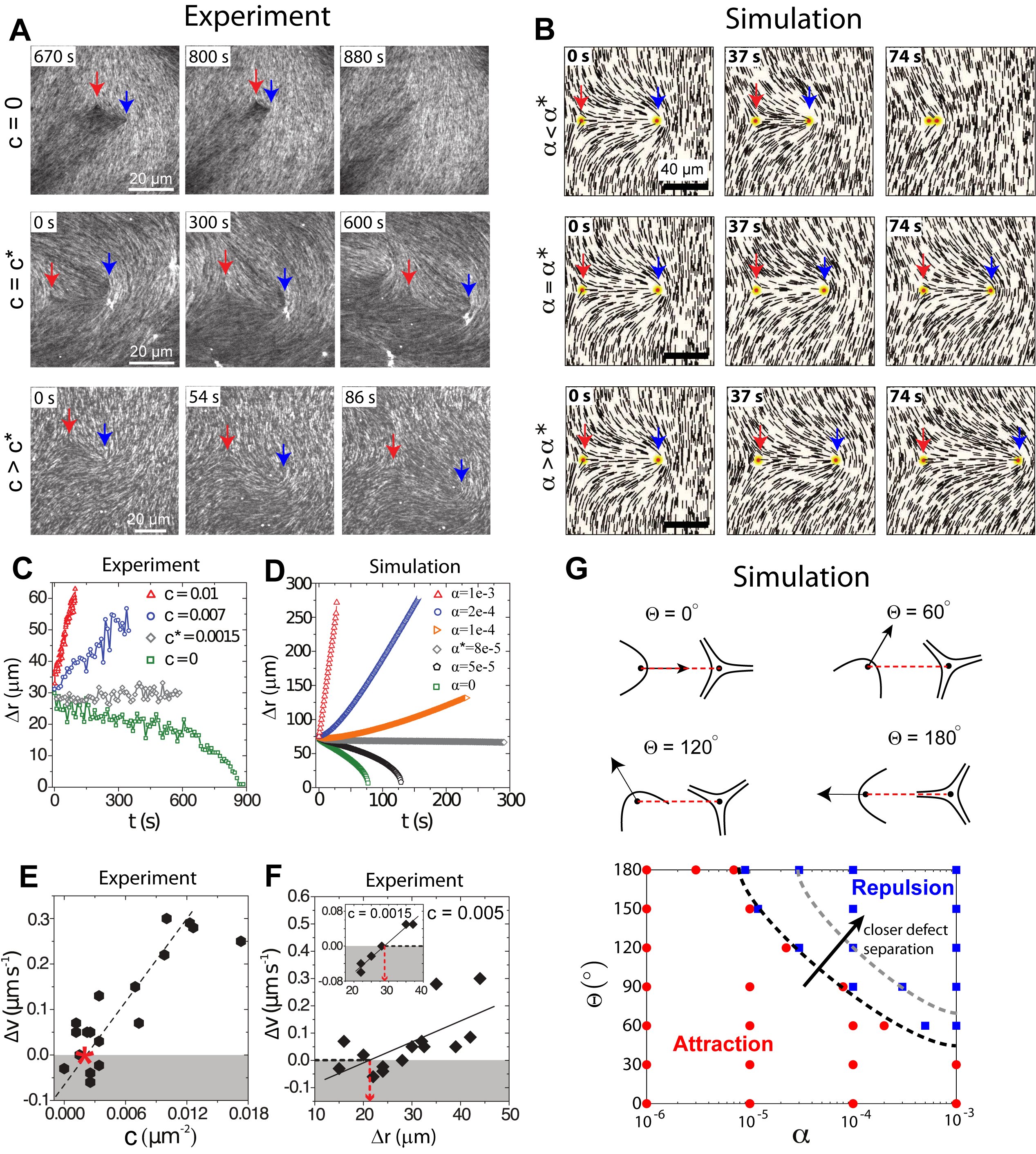}}
	\caption{\textbf{ Regulation of Defect Interactions by internal stress}. (A) Fluorescence images of actin LC showing dynamics of $\pm$1/2 defect pair (blue and red arrows, respectively) for varying levels of $c$ showing annihilation (top), stalling (middle) and repulsion (bottom). (B) The director field obtained around a $\pm$1/2 defect pair from the simulations at varying levels of active stress showing annihilation (top), stalling (middle) and repulsion (bottom) (C) Defect separation, $\Delta r$, as a function of time at different values of $c$ obtained from experimental data for defects with an initial separation of 30 $\mu$m. At $c^{*}$ = 0.00015 $\mu m^{-2}$, the defects spacing remains constant. (D) Defect separation, $\Delta r$, as a function of time as $\alpha$ is increased from 0 to 0.001 obtained from simulation data. (E) Relative velocity of defect separation, $\Delta v$, as function of $c$ obtained from experiments; the red asterisk corresponds to $c^{*}$. Dashed line is the linear fit to the data. (F) $\Delta v$ as a function $\Delta r$ for $c$ = 0.005 $\mu$m$^{-2}$ and 0.0015 $\mu$m$^{-2}$(inset). Solid black lines show linear fits. Red dashed line indicates the length scale where $\Delta v$ is zero. Data correspond to initial defect spacing, $\Delta r = 30~\mu$m. (G) Phase diagram of defect-pair dynamics in terms of activity $\alpha$ and initial relative orientation $\Theta$; the dashed line indicates that phase boundary moves when the defect separation becomes smaller.}
	\label{fig3}
\end{figure*}

Beyond LC structure, activity also influences dynamics.  In the absence of active stress ($c$ = 0 $\mu$m$^{-2}$), defects of opposite charge experience an attractive interaction, as the elastic energy is reduced through annihilation of +1/2 and -1/2 defects (Fig. \ref{fig3}A, top and Video 6). We quantify defect annihilation by tracking the distance between defect pairs, $\Delta r$, over time (Fig. \ref{fig3}C). Defect annihilation occurs slowly, at rates of 2 $\mu$m/min (Fig. \ref{fig3}C, green squares), a phenomenon that has been studied previously  \cite{toth2002hydrodynamics,RZhang}.  In contrast, at high motor density ($c>$ 0.0015 $\mu$m$^{-2}$), we observe that +1/2 and -1/2 defects effectively ``repel'' each other (Fig. \ref{fig3}A, bottom) and Video 7), such that the defect spacing increases at rates of $\ge$10 $\mu$m/min (Fig. \ref{fig3}C, red triangles and blue circles). A similar phenomenon, namely the ``unbinding of defects'', has been reported in microtubule-kinesin based systems \cite{dogic_nature2012} and 2D hydrodynamic simulations \cite{giomi_prl2013}. Here we examine this effect using our 3D simulations. Due to symmetry breaking in the surrounding director field, a $+1/2$ defect moves along its orientation (indicated by arrow in Fig.~\ref{fig1}D), activated by extensile stresses. In the absence of any far-field flows and elastic forces, simulations indicate that $+1/2$ defects are mobile while $-1/2$ defects remain relatively immobile. The transition from attractive to repulsive interaction between defects of opposite topological charge is also observed in the simulations in the range from $\alpha=0$ to $0.001$ (Fig. \ref{fig3}B+D), and can be qualitatively understood as activity generating propulsive stresses within the nematic field that are sufficiently strong to overcome elastic stresses.

Our simulations also indicate that a critical activity ($\alpha^*$) exists for which the propulsive stresses are perfectly balanced by the elasticity of the LC, leading to the ``stalling'' of defect pairs where their separation stays constant for several hundred seconds (Fig. \ref{fig3}B+D). We also observe such defect stalling experimentally at a critical motor density $c^*$ (Fig. \ref{fig3}A+C, Video 8). Both experiments and simulations show that although the inter-defect distance remains constant over the course of several hundred seconds, their positions shift over time, possibly due to uncontrolled background flows. This demonstrates that propulsive stresses from activity can be used to qualitatively alter the defect dynamics.

To quantify the change from attractive to repulsive behavior, we plot the relative speeds between paired defects ($\Delta v = v_{+1/2}-v_{-1/2}$) as a function of $c$ for our experimental data (Fig. \ref{fig3}E). This shows the transition from attractive to repulsive interactions for defects with an initial separation $\Delta r = 30~\mu$m occurs around $c$ = 0.003 $\mu$m$^{-2}$ and the relative velocity is linearly controlled by motor concentration. Finally, we find that, for a constant activity, $\Delta v$ also scales linearly with the initial defect separation, $\Delta r$ (Fig. \ref{fig3}F), such that we can define a length scale at which the transition between attractive and repulsive interactions occur.  Interestingly, we see evidence that this length scale increases from 20 $\mu$m to 30 $\mu$m as the motor density decreases from 0.005 $\mu$m$^{-2}$ to 0.0015 $\mu$m$^{-2}$. Understanding how activity can alter the nature of defect interactions over varying length scales will be an exciting topic for future research.

To generalize the above findings, we also consider arbitrary relative orientations of a defect pair, as illustrated in Fig.~\ref{fig3}G and SI Fig. \ref{s6}. The angle $\Theta$ between the $+1/2$ defect orientation and the line connecting the two defect cores has a profound effect on defect dynamics \cite{vromans_softmatt2016,tang_softmatt2017}. Using simulations (see Supplementary Text in SI for details), we explore how defect pair interactions are impacted by changes to $\Theta$ and activity $\alpha$ (Fig.~\ref{fig3}G). When $\Theta$ is small, as the $+1/2$ defect faces the $-1/2$ defect, their interaction is always attractive; when $\Theta$ is large, as the $+1/2$ defect points away from the $-1/2$ defect, there is a transition activity $\alpha^*(\Theta)$ (as a function of $\Theta$) above which defects become repulsive. Our simulations also show that when defect separation is closer, the phase boundary shifts to higher $\alpha$, a feature consistent with experimental observations (Fig.~\ref{fig3} E\&F). Thus, internal stresses can qualitatively change the interactions between defect pairs in LCs.

\subsection*{Defect density in an extensile active nematic is mainly determined by bend modulus}

\begin{figure*}
	\centerline{\includegraphics[width=0.8\textwidth]{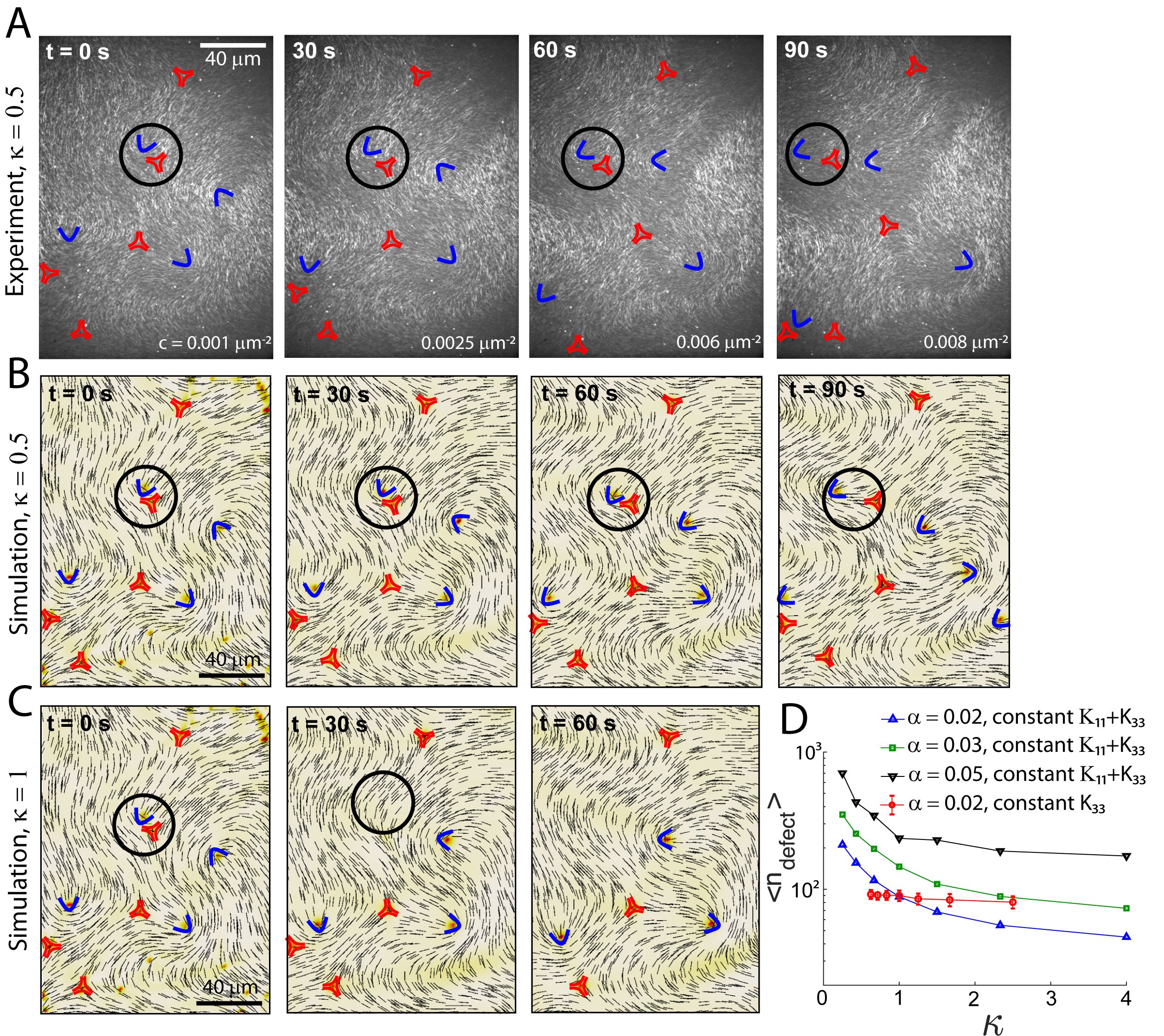}}
	\caption{\textbf{LC Mechanics is essential for predicting Active-state Dynamics and Structure}. (A) Time-lapse fluorescence images of an active actin-based LC ($\kappa$ = 0.5).  $\pm$1/2 defects are indicated by blue arcs and red triangles, respectively. The director field obtained from simulations of an active LC with the same instantaneous activity level as in the experiments, evolved over time. Simulation is initiated with the director field from the experiments in (A) at t = 0. The mechanics of the LC are $\kappa$ = 0.5 ($K_{33}$ = 0.5$K$, $K_{11}$ = $K$ with $K=1$~pN) in (B) and  $\kappa$ =1 ($K_{33}=K_{11} = 0.75K$) in (C).  The black circle highlights a defect pair that undergoes an annihilation event in (C) but not in (A) or (B). (D) Defect density as a function of $\kappa$. $<n_{defect}>$ decreases as a function of $\kappa$ when $K_{33}$ decreases, while keeping $K_{11}$ + $K_{33}$ constant for several different activity levels (black, green and blue symbols). In contrast, it merely changes when $K_{11}$ is varied while keeping $K_{33}$ constant (red symbols). }
	\label{fig4}
\end{figure*}

The inherent elasticity of a nematic LC can be viewed as a measure of the restoring force acting against spatial distortions \cite{deGennes_book}. In two dimensions, a nematic LC opposes splay ($K_{11}$) and bend ($K_{33}$) deformations but existing models of active nematics have been generally assumed $K_{11}=K_{33}$ \cite{RMP}. Our results in Fig. \ref{fig2} suggest this may be insufficient to faithfully capture active LC mechanical response and, thus, their dynamics. To explore this, we construct an LC comprised of actin filament length $l$ = 1 $\mu$m and use the +1/2 defect morphology to calculate $K_{33}/K_{11}$ = 0.5.  As described earlier, the addition of motors drives LC dynamics, as shown in the series of optical images in Fig. \ref{fig4}A. We design two simulation systems, one with $\kappa$ = 0.5 and another with $\kappa$ = 1, for which the initial director field is directly taken from the experiments at time t = 0 s (see Methods~\ref{m5}). The dynamics obtained from these simulations are shown in Fig. \ref{fig4}B and Fig. \ref{fig4}C respectively. We find that for $\kappa$ = 0.5, locations and trajectories of defects in Fig. \ref{fig4}B exhibit good agreement with experiments, whereas agreement is poor for $\kappa$ = 1. In particular, we find that the encircled defect pair undergoes annihilation for $\kappa$ = 1, an event that is not observed in the experimental data nor the simulations with accurate mechanical properties. This shows that the defect-dynamics at mesoscopic length and time scales strongly depends on the choice of splay and bend elasticity in the model. In order to isolate the roles of $K_{11}$ and $K_{33}$, we run simulations on a larger system size with variable elasticity. We find that the defect density, $<n_{defect}>$, defined as the total number of defects per unit area, decreases with $\kappa$ with constant $K_{11}+K_{33}$ at all activity values, as shown in Fig. \ref{fig4}D. Furthermore, by keeping $K_{33}$ constant and varying $K_{11}$ alone, we find that the defect density merely changes over a wide range of $\kappa$. Thus, for extensile active nematics, defect density in the active state is mainly controlled by $K_{33}$, by regulating the propensity of defect-pairs to annihilate.


\subsection*{Topological defects exhibit liquid-like structure and preferred orientations}

\begin{figure*}
	\centerline{\includegraphics[width=0.75\textwidth]{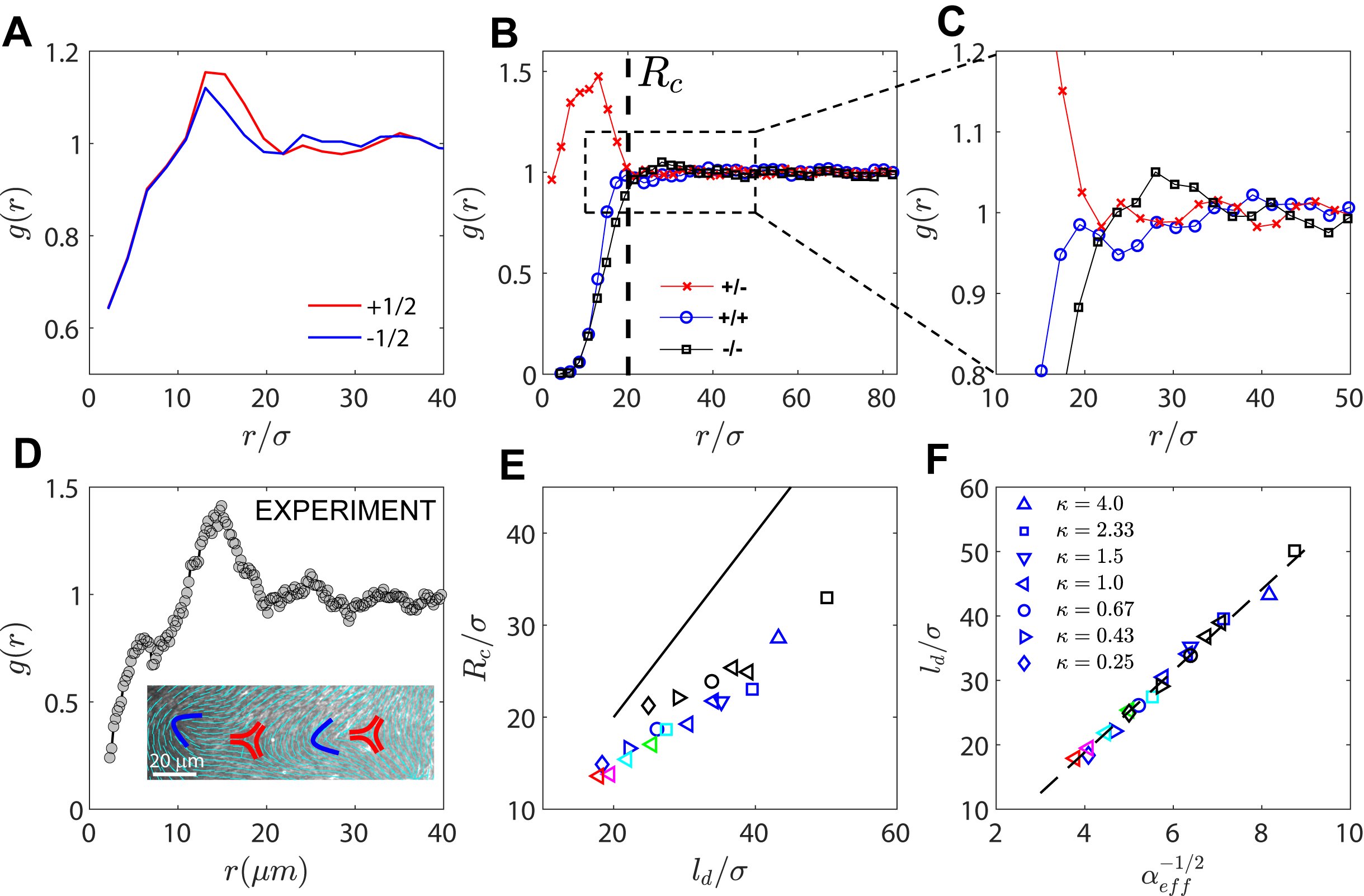}}
	\caption{\textbf{Radial Distribution Function of Defect Structure.} (A) Radial Distribution Function $g(r)$ for $+1/2$ and $-1/2$ defects from simulations for $\zeta=0.03$ and $\kappa=1.0$. (B) $g(r)$ of defects of specific charge ($+1/2$ and $-1/2$) from simulation. (C) Zoom-in of (B) reveals higher-order peaks in $g(r)$. (D) Radial Distribution Function $g(r)$ of topological defects from experiments with activity $c=0.005~\mu$m$^{-2}$. Inset shows experimental evidence of higher-order peaks in $g(r)$. (E) A characteristic length scale $R_c$ emerges from $g(r)$ (illustrated in (B)); it is plotted against average defect spacing $l_d$; Black line indicates $R_c=l_d$. (F) Average defect spacing $l_d$ is plotted against $\alpha_{eff}^{-1/2}$, where the effective activity is defined as $\alpha_{eff}=\alpha/\sqrt{\kappa}$.}
	\label{fig5}
\end{figure*}

To further understand the combined effects of activity and elasticity on the microstructure of active nematics, and gain insights into the seemingly chaotic behavior of topological defects, we rely on measures of order that have been particularly useful in the context of simple liquids, namely radial distribution functions ($g(r)$).
Note that the correlation length calculations presented in Fig.~\ref{fig1} neglect the presence of defects. As a complimentary analysis tool, we introduce $g(r)$ between defects and measure it as a function of activity. In this view, the active nematic system can be regarded as a binary system of positive and negative particles (defects) (Video 9).
In a first step, we ignore defect orientations and focus only on their spatial distribution. The radial distribution functions corresponding to defect cores in our active nematic system are akin to those observed in liquids, with a first peak corresponding to (+) and (-) defect pairs, and higher order peaks arising from longer-range correlations. The predictions of simulations (Fig.~\ref{fig5}A) are in semi-quantitative agreement with our experimental observations (Fig.~\ref{fig5}D). A shoulder is observed before the first peak of $g(r)$; it can be explained by inspecting the radial distribution functions corresponding to like-charge defects.

In Fig.~\ref{fig5}B we differentiate $+$ and $-$ defects when calculating $g(r)$ in simulation. We see a length scale $R_c$ exists below which the radial distribution of $\pm$ defects deviates considerably from unity. While $+/-$ defect pairs exhibit a pronounced peak at distances below $R_c$, like-charge defects exhibit short-range repulsions. The repulsive core therefore shows up as the shoulder in the total $g(r)$ seen in Fig. \ref{fig5} A\&D. By closely examining $g(r)$ at distances between 10 and 50 $\mu$m, we observe that $g(r)$ for $+/+$ defect pairs reaches a plateau earlier than that for the $-/-$ defect pair (Fig. \ref{fig2}C), implying that the average repulsive force between $+$ defect pairs is weaker than that between $-$ defect pairs. Higher order peaks in $g(r)$ at longer distances are clearly visible in $g(r)$, and can be explained by the fact that chains of alternating $\pm$1/2 defects are occasionally formed in these systems (Fig. \ref{fig5}D, inset). In Fig. \ref{fig5}E, we find that the emerging length scale $R_c$ exhibits a linear relation with the average defect spacing $l_d$. Thus, spacial inhomogeneity in defect charge becomes important when the defect separation is below the average spacing $l_d$.

These findings also imply that $l_d$ is a fundamental length scale that sets the system's defect structure. To examine the effect of elastic anisotropy, $l_d$ is plotted against an effective activity $\alpha_{eff}$, defined as $\alpha_{eff}=\alpha/\sqrt{\kappa}$. Fig.~\ref{fig5}F shows that all data collapse onto a master curve.
Because at rest ($0$ activity), systems of different $\kappa$ are degenerate, bearing the same $l_d=\infty$ at equilibrium, we say that elastic anisotropy modifies the system's activity, rather than that activity modifies elastic anisotropy.
Activity also breaks the symmetry of splay and bend. For the same activity level, extensile systems of lower $\kappa$ engender more defects than those of higher $\kappa$ with the same $K_{11}+K_{33}$. This is consistent with the microscopic view that extensile systems are unstable to bend instability $-$ low bend systems are prone to engender more defects.

\begin{figure*}
	\centerline{\includegraphics[width=0.7\textwidth]{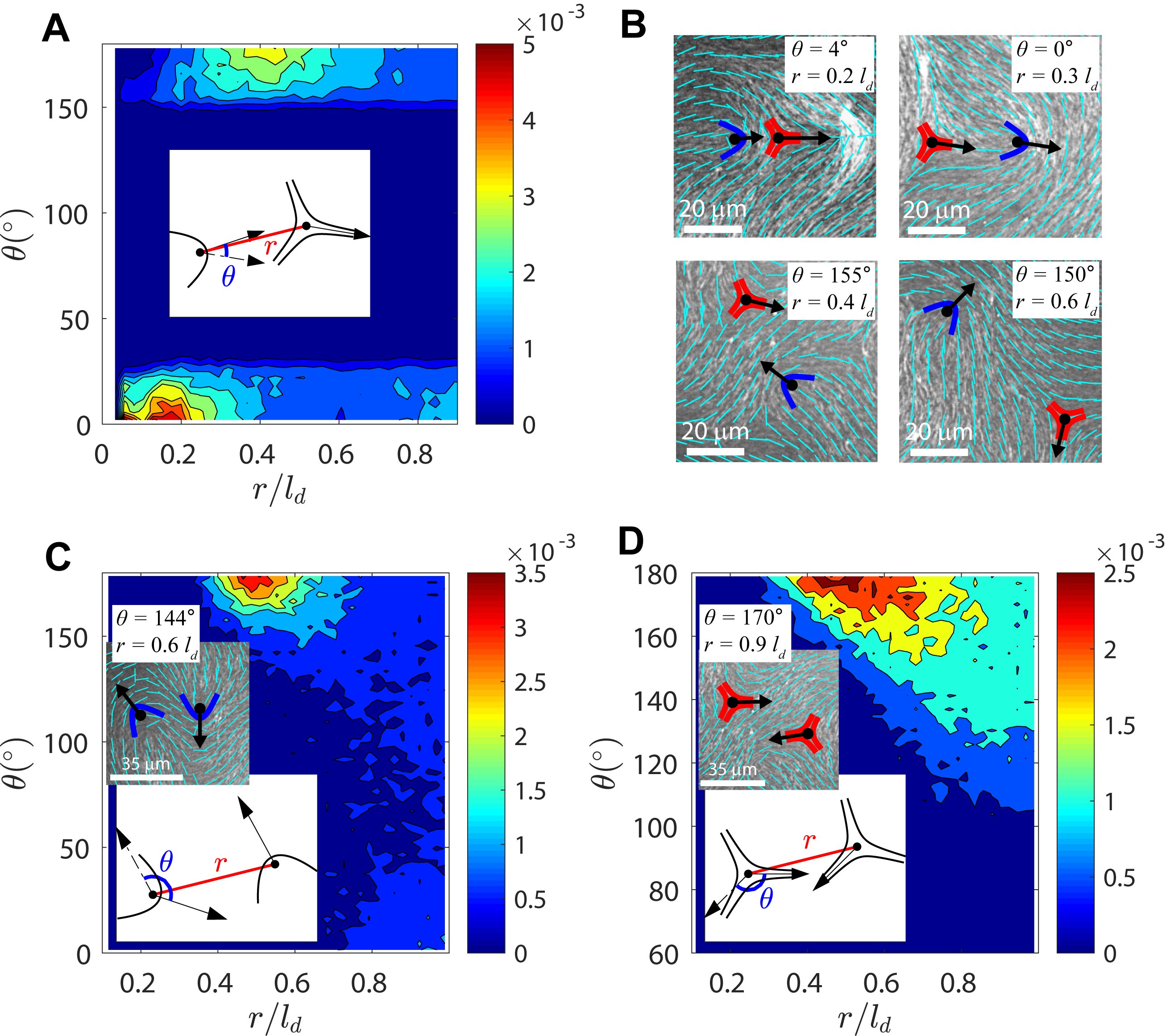}}
	\caption{\textbf{Analysis of Defect Orientational Structures.}. Probability distribution as function of defect distance $r$ and defect angle $\theta$ (schematically defined in inset plots) for $+/-$ (A), $+/+$ (C), and $-/-$ (D) defect pair. Inset images in C \& D show experimental observations of antiparallel like-charge defects (activity level $c=0.005~\mu$m$^{-2}$ and $l_{d}$ = 44 $\mu$m). (B) Typical structures of unlike-charge defect pairs observed in experiments. Top row: defect orientations tend to be parallel at short $r$; Bottom row: defect orientations tend to be antiparallel at intermediate $r$.}
	\label{fig6}
\end{figure*}

We next consider defect orientation, and study how it is coupled to defect separation. Fig.~\ref{fig6} shows three types of defect pair, namely are $+/-$ (Fig.~\ref{fig6}A), $+/+$ (Fig.~\ref{fig6}C), and $-/-$ defects (Fig.~\ref{fig6}D). By defining the angle between defect orientations, $\theta$, one can prepare a probability heat map in terms of $r$ and $\theta$. The definition of $+1/2$ defect orientation is illustrated in the insets of Fig.~\ref{fig6} and Fig. \ref{fig1}D.
Because a $-1/2$ defect has 3-fold symmetry, one has to choose one of its three branches to define its orientation. For a $+/-$ defect pair, we choose one branch as its orientation such that it is either parallel or antiparallel to the $+1/2$ defect's orientation (a minimizer of $\cos(|\theta|)$). For $-/-$ defect pairs, we choose one such that it makes the smallest angle with the defect position vector ${\bf r}$ (always pointing away from the defect of interest). We observe that for opposite-charge defect pairs, defects tend to align with each other when they are close (see Fig.~\ref{fig6}B for experimental images and Video 10). There are two equally possible scenarios in a steady-state system; in one, the $+1/2$ defect points towards the $-1/2$ defect (pre-annihilation event), and in the other, the $+1/2$ defect points away from the $-1/2$ defect (post-proliferation event).

Surprisingly, we find that there is a second stable regime for which $\pm 1/2$ defects are antiparallel at slightly longer separations $r$. This indicates that when defect spacing is in some intermediate range, the far field dictated by the $-1/2$ defect aligns the $+1/2$ defect in an antiparallel fashion. We have found abundant experimental evidences, some of which are shown in Fig.~\ref{fig6}B and Video 10, in support of this prediction. For like-charge pairs (see Fig.~\ref{fig6} C\&D), however, there is only one stable regime in which defects are antiparallel (face-to-face or back-to-back) to each other.
Note that the above calculations are similar to Fig.~\ref{fig3}G in terms of understanding defect orientations, but they are addressing different physics. In Fig.~\ref{fig3}G, we examined the dynamics of an isolated defect pair at low activity, when the active stress is balanced by the elastic forces arising from existing defects. In contrast, in Fig.~\ref{fig6}, we collected statistics for hundreds of interacting defect pairs in the high activity regime, where activity is dissipated by generating new defects.
Taken together, our findings indicate that defects in active systems can be described in terms of liquid state correlations, and that their interactions are anisotropic, with an interesting angular dependence that could potentially be used to engineer intricate transport mechanisms within these active materials.

\section*{Discussion}

Our work demonstrates the emergence of an active nematic in actin-based LC driven by myosin-II motors. This system closely resembles the active nematics that are formed by microtubule filaments and kinesin motors \cite{dogic_nature2012}. One striking finding is that active nematics can be realized with punctate myosin filaments, which contain $\sim$ 100s of motor heads and are sparsely distributed. While the kinesin tetramers used to realize active microtubule-based nematics have not been directly visualized, we presume they would be more homogeneously distributed across the nematic. That stress inhomogeneities do not negatively impact the realization of active nematics underscores that these systems are dominated by long-range hydrodynamic and elastic effects. In both systems, motor-filament interactions give rise to uniaxial extensile stress. Previous work has shown that contractile stress dominates in actomyosin systems as filament length increases \cite{SamPNAS} or with the addition of cross-linking proteins \cite{MikeMurrell}. Further work will be needed to map out how the force generation by motor-filament interactions can be tuned by filament length, stiffness, and cross-linking \cite{SamPNAS}.

The similarities between actin and microtubule systems not withstanding, there are several quantitative differences that should be noted. First, activity-induced changes in defect shape have not been reported in microtubule-based nematics. We expect that higher levels of activity may be needed to overcome the higher rigidity of microtubules, which are 1000-fold stiffer than actin filaments. Furthermore, another significant difference between the two systems lies in the steady-state defect structure. In actin-myosin nematics, $g(r)$ shows higher order peaks, which indicate a strong interaction between defects; this more pronounced structure has not been reported in microtubule-kinesin experiments \cite{dogic_natmatt2015}. A possible explanation for this difference might be the defect density, which is 5-fold higher in actin than reported for microtubule-kinesin nematics.


In summary, we have performed experiments and simulations on a quasi-2D active nematic liquid crystal composed of short actin filaments driven by myosin motors. The clustering dynamics of myosin II motors have allowed us to investigate how the structure and dynamics of liquid crystals vary as a function of internal activity. We characterize the  motor driven changes in structure and flows that arise in terms of the characteristic correlation lengths and defect density as a function of motor density, and find dependencies that are fully captured by non-equilibrium hydrodynamic simulations. Our combined theoretical and experimental approach has allowed us to use the change in +1/2 defect morphology induced by activity to reveal the change in effective bend elasticity resulting from the microscopic stresses.  We demonstrate that the activity can fundamentally change the nature of defect-pair interactions, from attractive to effective repulsions, and we show that it is possible to control the relative defect speeds with motor concentration. The critical activity is shown to be a function of initial defect separation and relative orientation.
Our further calculations of correlations of defects show that their configurations exhibit liquid-like structure, and that the relative orientations of defect pairs become highly correlated when they are in close proximity.

\subsection*{Acknowledgments}

This work was supported primarily by the University of Chicago Materials Research Science and Engineering Center, which is funded by the National Science Foundation under award number DMR-1420709. MLG acknowledges support from the ARO MURI Grant W911NF1410403.  JJdP acknowledges support from NSF Grant DMR-1710318. NK acknowledges the Yen Fellowship of the Institute for Biophysical Dynamics, The University of Chicago. The authors thank Dr. Kimberly Weirich for useful discussions and purified proteins, Dr. Samantha Stam for assisting with experiments, and Dr. Patrick Oakes for helping in director field analysis. RZ is grateful for the support of the University of Chicago Research Computing Center for assistance with the calculations carried out in this work. 

\textbf{Author Contributions:} N.K., R.Z., J.J.d.P., and M.L.G. designed research; N.K. performed experiments and R.Z. carried out simulations; N.K. and R.Z. analyzed data; and N.K., R.Z., J.J.d.P., and M.L.G. wrote the paper.

\section*{Methods}
\subsection{Experimental methods}
\subsubsection{Proteins}
Monomeric actin is purified from rabbit skeletal muscle acetone powder \cite{actin} (Pel-Freez Biologicals, Rogers, AR) and is stored at $-80^0$C in G-buffer(2mM Tris HCL pH 8.0, 0.2 mM APT, 0.2 mM CaCl$_{2}$, 0.2 mM DTT, $0.005\%$ NaN$_{3}$). Tetramethylrhodamine-6-maleimide dte (Life Technologies, Carlsbad, CA) is used to label actin. Capping protein [mouse, with a HisTag, purified from bacteria \cite{CP}; gift from the Dave Kovar lab, The University of Chicago, Chicago, IL] is used to regulate actin polymerization and shorten the filament length. Skeletal muscle myosin II is purified from chicken breast \cite{Myo1} and labeled with Alexa-642 maleimide (Life Technologies, Carlsbad, CA) \cite{Myo2}.

\subsubsection{Experimental assay and microscopy}
The actin is polymerized in 1X F-buffer (10 mM imidazole, pH 7.5, 50mM KCL, 0.2mM EGTA, 1mM MgCl$_{2}$ and 1mM ATP). To avoid photobleaching, an oxygen scavenging system (4.5 mg/mL glucose, 2.7 mg/mL glucose oxidase(cat$\#$345486, Calbiochem, Billerica, MA), 17000 units/mL catalase (cat $\#$02071, Sigma, St. Louis, MO) and 0.5 vol. $\%$ $\beta$-mercaptaethanol is added. 0.3 wt $\%$ 15 cP methylcellulose is used as the crowding agent. Actin from frozen stocks stored in G-buffer is added to a final concentration of 2 $\mu$M with a ratio 1:5 TMR-maleimide labeled:unlabeled actin monomer. Frozen capping protein stocks are thawed on ice and are added at the same time (6.7 and 3.3 nM for 1 $\mu$m and 2 $\mu$m long actin filaments). We call this assay ``polymerization mixture'' from henceforth. Myosin-II is mixed with phalloidin-stabilized F-actin at a 1:4 myosin/actin molar ratio in spin-down buffer (20 mM MOPS, 500 mM KCL, 4mM MgCl2, 0.1mM EGTA; pH 7.4) and centrifuged for 30 min at 100,000 g. The supernatant containing myosin with low affinity to F-actin is used in experiments whereas the high-affinity myosin is discarded.

Experiment is performed in a glass cylinder (cat$\#$ 09-552-22, Corning Inc.) glued to a coverslip \cite{SamPNAS}. Coverslips are cleaned by sonicating in water and ethanol. The surface is treated with triethoxy(octyl)silane in isopropanol to produce a hydrophobic surface. To prepare a stable oil-water interface, PFPE-PEG-PFPE surfactant (cat $\#$ 008, RAN biotechnologies, Beverly, MA) is dissolved in Novec-7500 Engineered Fluid (3M, St Paul, MN) to a concentration of 2 wt. $\%$. To prevent flows at the surface, a small $2\times2$ mm teflon mask is placed on the treated coverslip before exposing it to UV/ozone for 10 minutes. The glass cylinder is thoroughly cleaned with water and ethanol before gluing it to the coverslip using instant epoxy. Then $3~\mu$L of oil-surfactant solution is added into the chamber, and quickly pipetted out to leave a thin coating. The sample is always imaged in the middle of the film over the camera field of view, which is approximately 200 $\mu$m $\times$ 250 $\mu$m to make sure that the sample remains in focus over this area, which is far away from the edges. Imaging close to the edges is avoided. The polymerization mixture is immediately added afterwards. 30$-$60 minutes later, a thin layer of actin LC is formed. Myosin II motors are added to the polymerization mixture at 5$-$10 nM concentration.

The sample is imaged using an inverted microscope (Eclipse Ti-E; Nikon, Melville, NY) with a spinning disk confocal head (CSU-X, Yokagawa Electric, Musashino, Tokyo, Japan), equipped with a CMOS camera (Zyla-4.2 USB 3; Andor, Belfast, UK). A 40X 1.15 NA water-immersion objective (Apo LWD; Nikon) was used for imaging. Images were collected using 568 nm and 642 nm excitation for actin and myosin respectively. Image acquisition was controlled by Metamorph (Molecular Devices, Sunnyvale, CA).

\subsubsection{Image and data analysis}
The nematic director field is extracted the same way as in \cite{RZhang} which utilizes an algorithm which is described in detail in methods of Cetera et al. \cite{patrick}. In short this is as follows: the optical images were bandpassed filtered and unsharp masked in the ImageJ software \cite{imagej} to remove noise and spatial irregularities in brightness. The image algorithm computes 2D FFT of a small local square sections (of side $\psi$) of the image and uses an orthogonal vector to calculate the local actin orientation. The sections were overlapped over a distance $\zeta$ to improve statistics. $\psi$ and $\zeta$ are varied over 15-30 $\mu$m and 1-3 $\mu$m respectively for different images to minimize errors in the local director without changing the final director field.

Myosin puncta density was calculated using ImageJ software. Towards the end of the experiment, large clusters of myosin are not counted. Since the number of myosin polymers remains at least 10-fold greater than that of myosin clusters, our results are insensitive to the choice of the cluster cutoff size. We calculate the mean $l_d$, $\xi_\theta$ and $\xi_v$ from overlapping 150 s intervals. We explored averaging over shorter time intervals, and found that the trend in $l_d$ was similar but, as expected, the standard deviation increased (SI Fig. \ref{s4}). At the fastest rates of decrease, the myosin density does not decrease over this interval but is within the measurement error reported in Fig. \ref{fig1}C. The typical relaxation time of the actin nematic LC is given by $\tau_{R} = \gamma l^2/K$, where $\gamma$, $l$ and $K$ are the rotational viscosity, the filament length and the LC elastic modulus, respectively. For $\gamma \sim 0.1$ Pa.s, $l = 1$ $\mu$ m and $K$ = 0.13 pN, we find that $\tau_{R} \sim 1$ s. Thus, the LC structure achieves steady state on time scales much faster than the evolution of the myosin density.

The active flows are quantified using particle image velocimetry (available at \url{www.oceanwave.jp/softwares/mpiv/}) to extract local velocity field, $\textbf{v}$. The orientational correlation length, $\xi_\theta$ was calculated by computing $\int dr \frac{g_2(r)}{g(0)}$ where $g_2(r) = \langle\cos[2(\theta_i-\theta_j)]\rangle$ indicating spatial pairs $i$ and $j$ separated by a distance of $r$. Similarly, $\xi_v=\int dr \frac{\langle{\bf v}_i(0)\cdot {\bf v}_j(r)\rangle}{\langle v^2\rangle}$.

\subsection{Theory and modeling}
\subsubsection{Theoretical Model}
The bulk free energy of the nematic LC, $F$, is defined as
\begin{equation}\label{total}
\begin{aligned}
F&= \int_{V} dV f_{bulk} + \int_{\partial V} dS f_{surf} \\
&= \int_{V} dV (f_{LdG}+f_{el}) + \int_{\partial V} dS f_{surf},
\end{aligned}
\end{equation}
where $f_{LdG}$ is the short-range free energy, $f_{el}$ is the long-range elastic energy, and $f_{surf}$ is the surface free energy due to anchoring. $f_{LdG}$ is given by a Landau-de Gennes expression of the form\cite{deGennes_book,landau_statistics_book}:
\begin{equation}\label{phase}
f_{LdG}= \frac{A_0}{2} (1-\frac{U}{3} )\tr({\bf Q}^2) - \frac{A_0U}{3} \tr({\bf Q}^3) + \frac{A_0 U}{4}(\tr({\bf Q}^2))^2.
\end{equation}
Parameter $U$ controls the magnitude of $q_0$, namely the equilibrium scalar order parameter via $q_0=\frac{1}{4}+\frac{3}{4}\sqrt{1-\frac{8}{3U}}$.
The elastic energy $f_{el}$ is written as ($Q_{ij,k}$ means $\partial_k Q_{ij}$):
\begin{equation}\label{elastic_en}
\begin{aligned}
f_{el}=&\frac{1}{2}L_1 Q_{ij,k}Q_{ij,k}+\frac{1}{2}L_2 Q_{jk,k}Q_{jl,l}\\
&+\frac{1}{2}L_3 Q_{ij}Q_{kl,i}Q_{kl,j}+\frac{1}{2}L_4 Q_{ik,l}Q_{jl,k}.
\end{aligned}
\end{equation}
If the system is uniaxial, the above equation is equivalent to the Frank Oseen expression:
\begin{equation}\label{frank}
\begin{aligned}
f_e= \frac{1}{2}K_{11} &(\nabla\cdot {\bf n})^2+\frac{1}{2}K_{22}({\bf n}\cdot \nabla \times{\bf n})^2+\frac{1}{2}K_{33}({\bf n}\times (\nabla \times{\bf n}))^2 \\
& -\frac{1}{2}K_{24}\nabla\cdot [{\bf n}(\nabla\cdot{\bf n})+{\bf n}\times(\nabla\times{\bf n})].
\end{aligned}
\end{equation}
The $L$'s in Eq.~\ref{elastic_en} can then be mapped to the $K$'s in Eq.~\ref{frank} via
\begin{equation}
\begin{aligned}
L_1&=\frac{1}{2q_0^2} \left[ K_{22}+\frac{1}{3}(K_{33}-K_{11})  \right], \\
L_2&=\frac{1}{q_0^2} (K_{11}-K_{24}), \\
L_3&=\frac{1}{2q_0^3} (K_{33}-K_{11}), \\
L_4&=\frac{1}{q_0^2} (K_{24}-K_{22}).
\end{aligned}
\end{equation}
By assuming a one-elastic-constant $K_{11}=K_{22}=K_{33}=K_{24}\equiv K$, one has $L_{1}=L\equiv K/2q_0^2$ and $L_{2}=L_{3}=L_{4}=0$.
Point wise, ${\bf n}$ is the eigenvector associated with the greatest eigenvalue of the ${\bf Q}$-tensor at each lattice point.

To simulate the LC's non-equilibrium dynamics, a hybrid lattice Boltzmann method is used to simultaneously solve a Beris-Edwards equation and a momentum equation which accounts for the hydrodynamic effects.
By introducing a velocity gradient $W_{ij}=\partial_j u_i$, strain rate $\bf A=(\bf W + \bf W^T)/2$, vorticity $\bf \Omega=(\bf W - \bf W^T)/2$, and a generalized advection term
\begin{equation}
\begin{aligned}
{\bf S}({\bf W},{\bf Q})=&(\xi {\bf A}+{\bf \Omega})({\bf Q}+{\bf I}/3)+({\bf Q}+{\bf I}/3)(\xi {\bf A}-{\bf \Omega})&\\
&-2\xi ({\bf Q}+{\bf I}/3) \tr({\bf QW}),&
\end{aligned}
\end{equation}
one can write the Beris-Edwards equation\cite{beris_edwards_book} according to
\begin{equation} \label{beris_edwards_eq}
(\partial_t +{\bf u}\cdot \nabla){\bf Q}-{\bf S}({\bf W},{\bf Q})=\Gamma \bf{H}.
\end{equation}
The constant $\xi$ is related to the material's aspect ratio, and $\Gamma$ is related to the rotational viscosity $\gamma_1$ of the system by $\Gamma=2q_0^2/\gamma_1$\cite{denniston_2d}. The molecular field $\bf{H}$, which drives the system towards thermodynamic equilibrium, is given by
\begin{equation}
{\bf H}=-\left[ \frac{\delta F}{ \delta \bf{Q}} \right]^{st},
\end{equation}
where $\left[ ...\right]^{st}$ is a symmetric and traceless operator. When velocity is absent, i.e. ${\bf u}({\bf r})\equiv 0$, Besris-Edwards equation Eq.~\ref{beris_edwards_eq} reduce to Ginzburg-Landau equation:
$$
\partial_t {\bf Q}=\Gamma {\bf H}.
$$
To calculate the static structures of $\pm 1/2$ defects, we adopt the above equation to solve for the ${\bf Q}$-tensor at equilibrium.

Degenerate planar anchoring is implemented through a Fournier-Galatola expression\cite{fournier_galatola} that penalizes out-of-plane distortions of the ${\bf Q}$ tensor. The associated free energy expression is given by
\begin{equation}
f_{surf} = W (\tilde{\bf Q} - \tilde{\bf Q}^{\perp})^2,
\end{equation}
where $\tilde{{\bf Q}} = {\bf Q} + (q_0/3)\bf{I}$ and $\tilde{\bf{Q}}^{\perp} = {\bf P} \tilde{\bf{Q}} {\bf P}$. Here ${\bf P}$ is the projection operator associated with the surface normal ${\bf \nu}$ as ${\bf P}=\bf{I} -{\bf \nu} {\bf \nu} $. The evolution of the surface ${\bf Q}$-field is governed by\cite{zhang_jcp2016}:
\begin{equation} \label{surface_evolution}
\frac{\partial {\bf Q}}{\partial t}=-\Gamma_s \left( -L {\bf \nu} \cdot \nabla {\bf Q} +\left[  \frac{\partial f_{surf}}{\partial {\bf Q}} \right]^{st} \right),
\end{equation}
where $\Gamma_s=\Gamma/\xi_N$ with $\xi_N=\sqrt{L_1/A_0}$, namely nematic coherence length.

Using an Einstein summation rule, the momentum equation for the nematics can be written as\cite{denniston_2d,denniston_3d}
\begin{equation}  \label{ns_eq}
\begin{aligned}
\rho(\partial_t+u_j \partial_j)u_i=&\partial_j \Pi_{i_j}+\eta\partial_j[\partial_i u_j+\partial_j u_i +(1-3\partial_\rho P_0)\partial_\gamma u_\gamma \delta_{i_j}].
\end{aligned}
\end{equation}
The stress ${\bf \Pi}={\bf \Pi}^{p}+{\bf \Pi}^{a}$ consists of a passive and an active part. The passive stress ${\bf \Pi}^{p}$ is defined as
\begin{equation}
\begin{aligned}\label{stress}
\Pi_{i j}^{p}=  & -P_0 \delta_{i j}-\xi H_{i \gamma} (Q_{\gamma j} +\frac{1}{3}\delta_{\gamma j}) - \xi (Q_{i \gamma} +\frac{1}{3} \delta_{\gamma j} ) H_{\gamma j} & \\
 &+ 2 \xi (Q_{i j} +\frac{1}{3}\delta_{i j}) Q_{\gamma \epsilon}H_{\gamma \epsilon} -\partial_{j} Q_{\gamma \epsilon} \frac{\delta \mathcal F}{\delta \partial_i Q_{\gamma \epsilon}} &\\
 & + Q_{i \gamma} H_{\gamma j} -H_{i \gamma} Q_{\gamma j}, &
\end{aligned}
\end{equation}
where $\eta$ is the isotropic viscosity, and the hydrostatic pressure $P_0$ is given by\cite{fukuda_force}
\begin{equation}
P_0=\rho T - f _{bulk}.
\end{equation}
The temperature $T$ is related to the speed of sound $c_s$ by $T=c_s^2$. The active stress reads\cite{yeomans_pre2007}
\begin{equation} \label{active_stress}
\Pi_{i j}^{a}=  -\alpha Q_{i j},
\end{equation}
in which $\alpha$ is the activity in the simulation. The stress becomes extensile when $\alpha>0$ and contractile when $\alpha<0$.

\subsubsection{Numerical Details}\label{m5}
We solve the evolution equation Eq.~\ref{beris_edwards_eq} using a finite-difference method. The momentum equation Eq.~\ref{ns_eq} is solved simultaneously via a lattice Boltzmann method over a D3Q15 grid\cite{zhaoliguo_book}. The implementation of stress follows the approach proposed by Guo {\it et al.}\cite{zhaoliguo_forcing}. The units are chosen as follows: the unit length $a$ is chosen to be $a=\xi_N=1~\mu m$, characteristic of the filament length, the characteristic viscosity is set to $\gamma_1$=0.1~Pa$\cdot$s, and the force scale is made to be $F_0=10^{-11}$~N. Other parameters are chosen to be $A_0=0.1$, $K=0.1$, $\xi=0.8$, $\Gamma=0.13$, $\eta=0.33$, and $U=3.5$ leading to $q_0\approx 0.62$.
The simulation is performed in a rectangular box. The boundary conditions in the $xy$ plane are periodic with size $[N_x, N_y]=[250,~250]$. Two confining walls are introduced in the $z$ dimension, with strong degenerate planar anchoring, ensuring a quasi 2D system with z-dimension $7\le N_z\le11$.
We refer the reader to Ref.~\cite{zhang_jcp2016} for additional details on the numerical methods employed here.

\bibliography{ActiveCombined}
\newpage

\section*{Supplementary Information}

\subsection*{Supplementary Figures}

\begin{figure*}[h]
	\centerline{\includegraphics[width=0.55\textwidth]{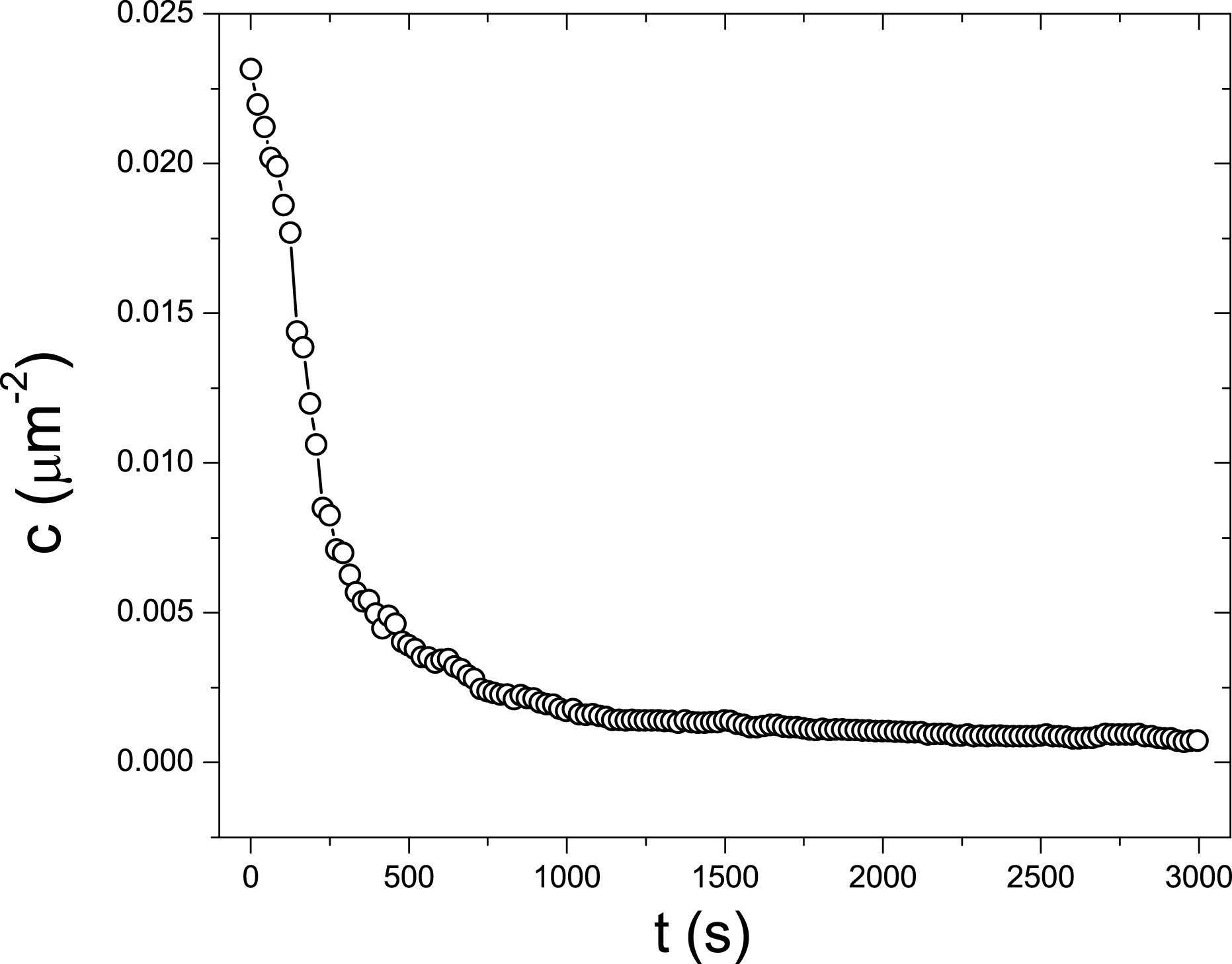}}
	\caption{\textbf{Myosin motors cluster over time.} The number density of motors $c$ decays as a function of time.}
	\label{s1}
\end{figure*}

\begin{figure}[h]
	\centerline{\includegraphics[width=0.55\textwidth]{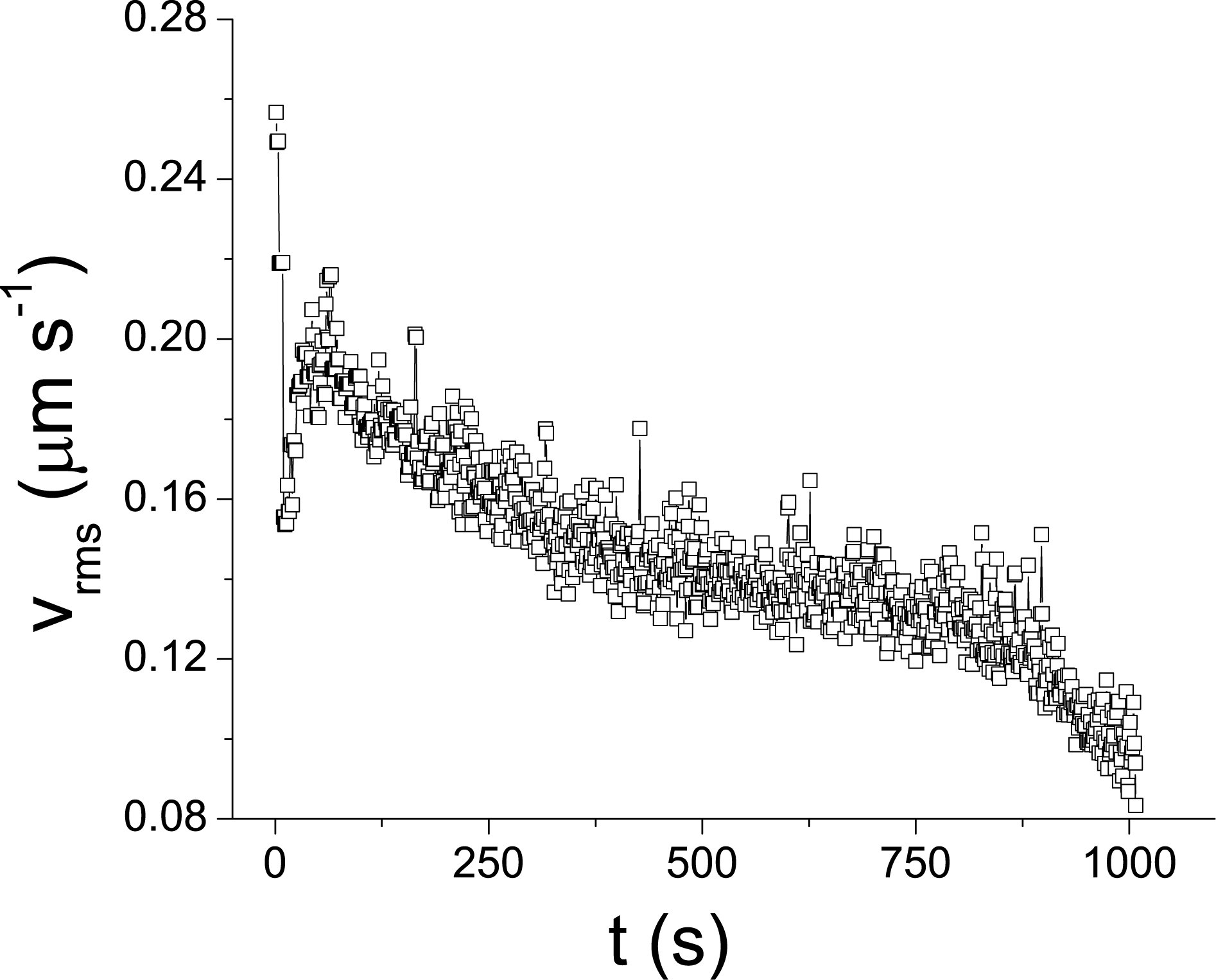}}
	\caption{\textbf{Temporal behavior of root mean square velocity.} Root mean square velocity ($v_{rms}$ = $\sqrt {\langle v^2 \rangle}$, where $v$ is the magnitude of local PIV velocity field) of the active nematic as a function of time. The figure shows that the kinematic energy is declining as motor proteins aggregate.}
	\label{s2}
\end{figure}
\newpage

\begin{figure}[h]
	\centerline{\includegraphics[width=.7\textwidth]{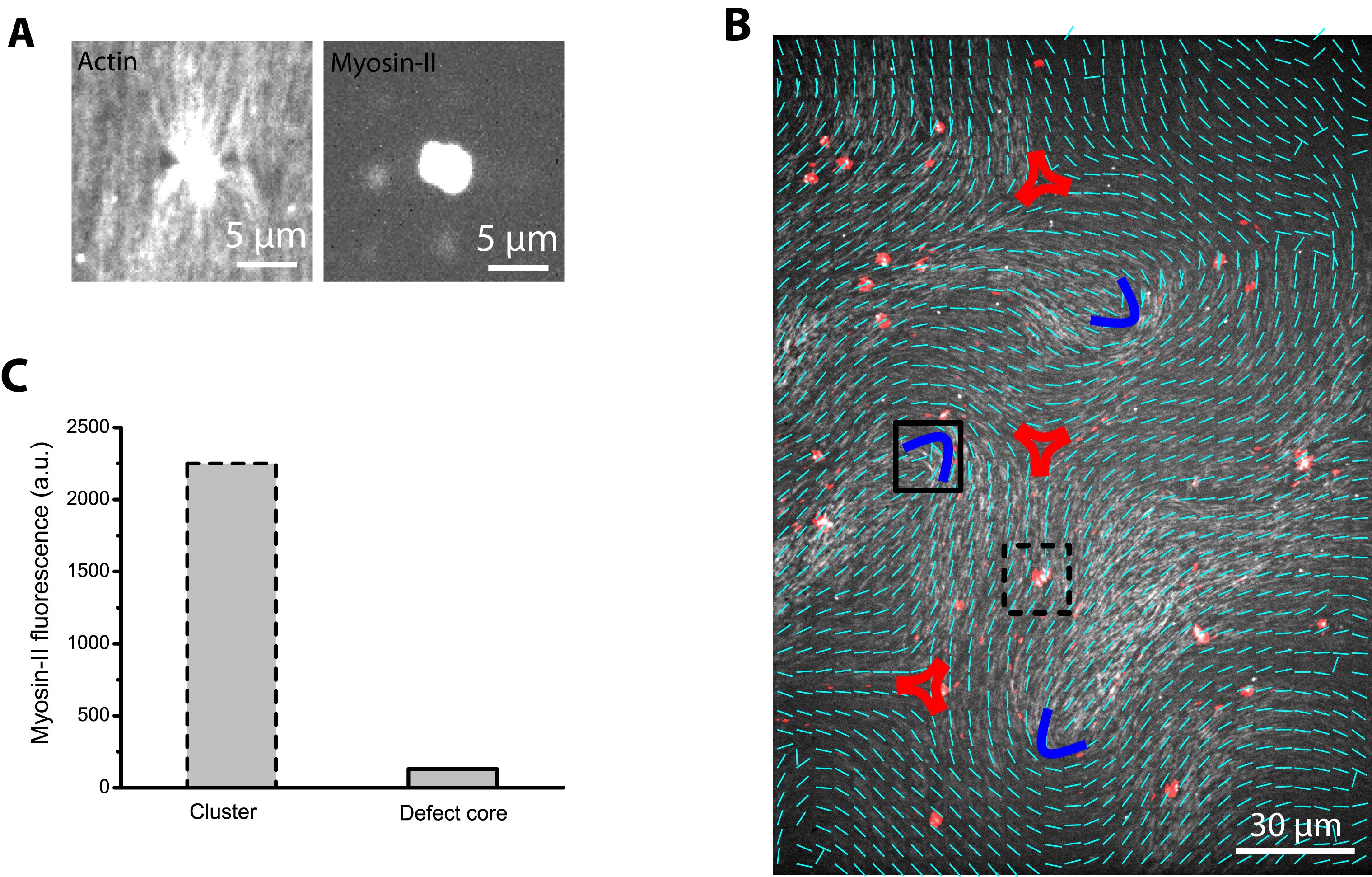}}
	\caption{\textbf{Myosin motors do not localize to defect cores.} (A) Actin and myosin-II channels corresponding to a typical cluster. (B) Experimental image with nematic director field overlaid, showing that the myosin clusters (in red) never occupy the defect cores. The +1/2 and -1/2 defects are indicated by blue and green symbols, respectively. (C) Average of the maximum value of myosin-II fluorescent signal at the location of a cluster (black dashed box) and close to the defect (black box).}
	\label{s3}
\end{figure}


\begin{figure}[h]
	\centerline{\includegraphics[width=0.45\textwidth]{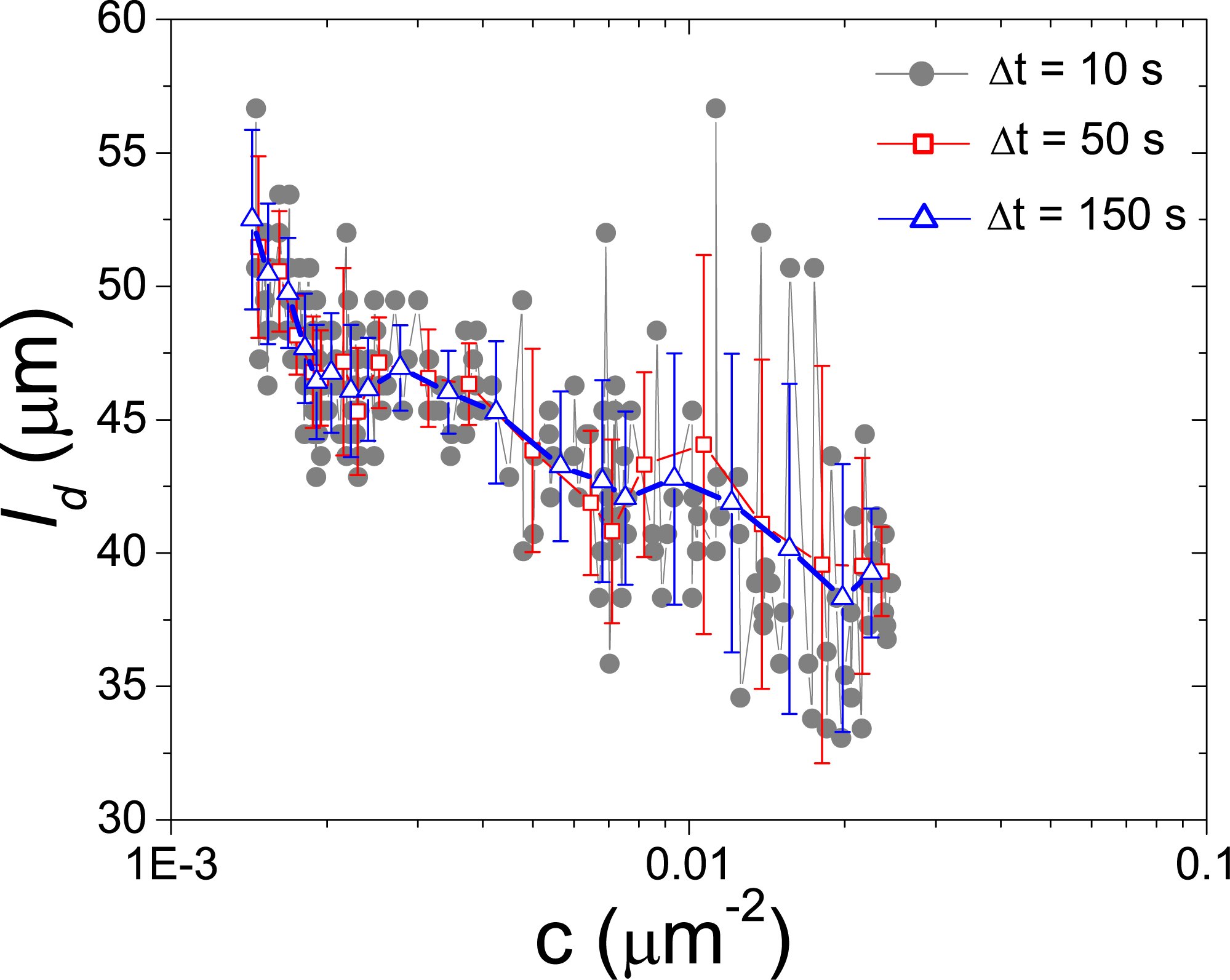}}
	\caption{\textbf{Time averaging of defect spacing.} Mean defect spacing, $l_d$, as a function of concentration for three values of time averaging time-interval. The data presented in main Fig. \ref{fig1}C correspond to a time-averaged data over a time window, $\Delta$t = 150 s, which corresponds to the minimum standard deviation without losing the general trend. The value $\Delta$t = 150 s is also used for $\xi_v$ and $\xi_\theta$. These timescales are much longer than the average relaxation time ($\sim$ 1 s) of the LC. }
	\label{s4}
\end{figure}


\begin{figure}[h]
	\centerline{\includegraphics[width=0.8\textwidth]{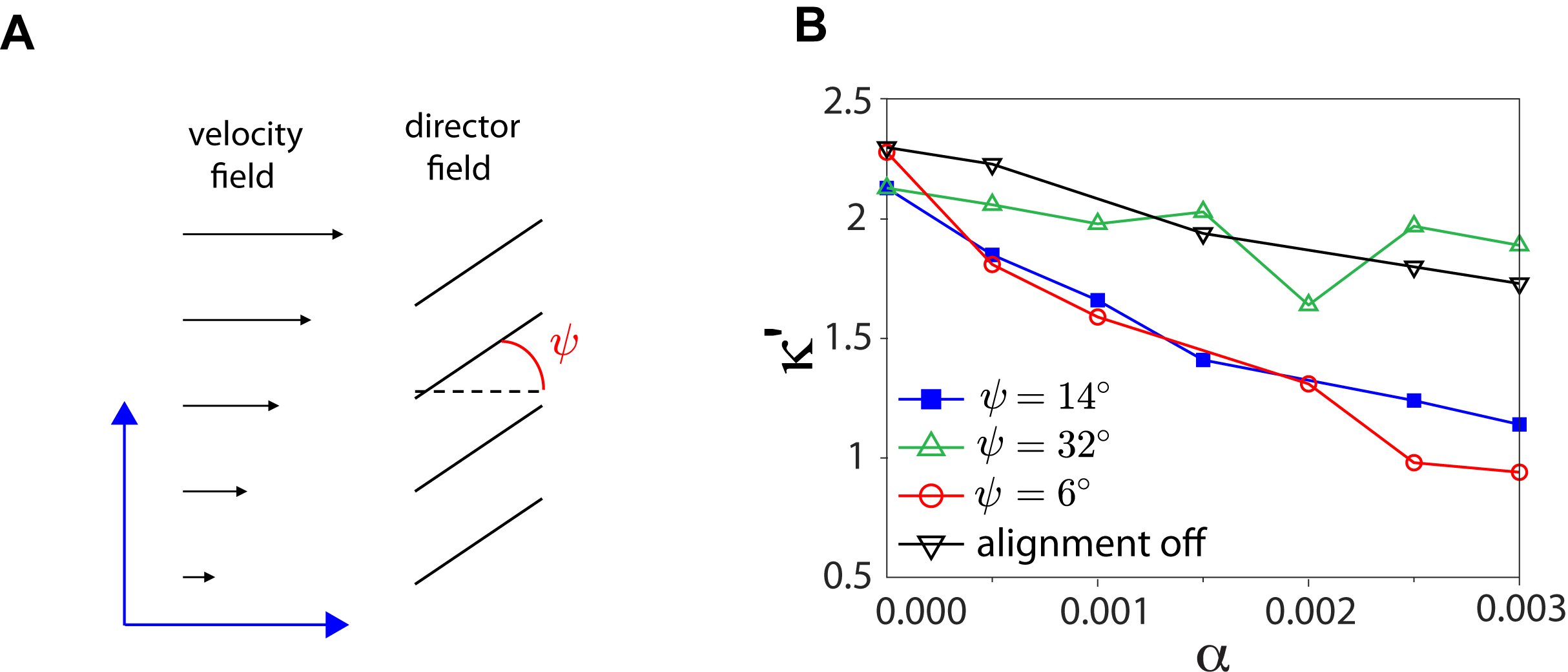}}
	\caption{{\bf Effect of flow-alignment on the change of defect morphology.} (A) Illustration of Leslie angle $\psi$ of liquid crystal subject to strong shear flow. (B) Apparent elasticity $\kappa'$ as function of activity $\alpha$ at different Leslie angles and when flow alignment is turned on or off. }
	\label{s5}
\end{figure}

\newpage

\begin{figure}[h]
	\centerline{\includegraphics[width=0.6\textwidth]{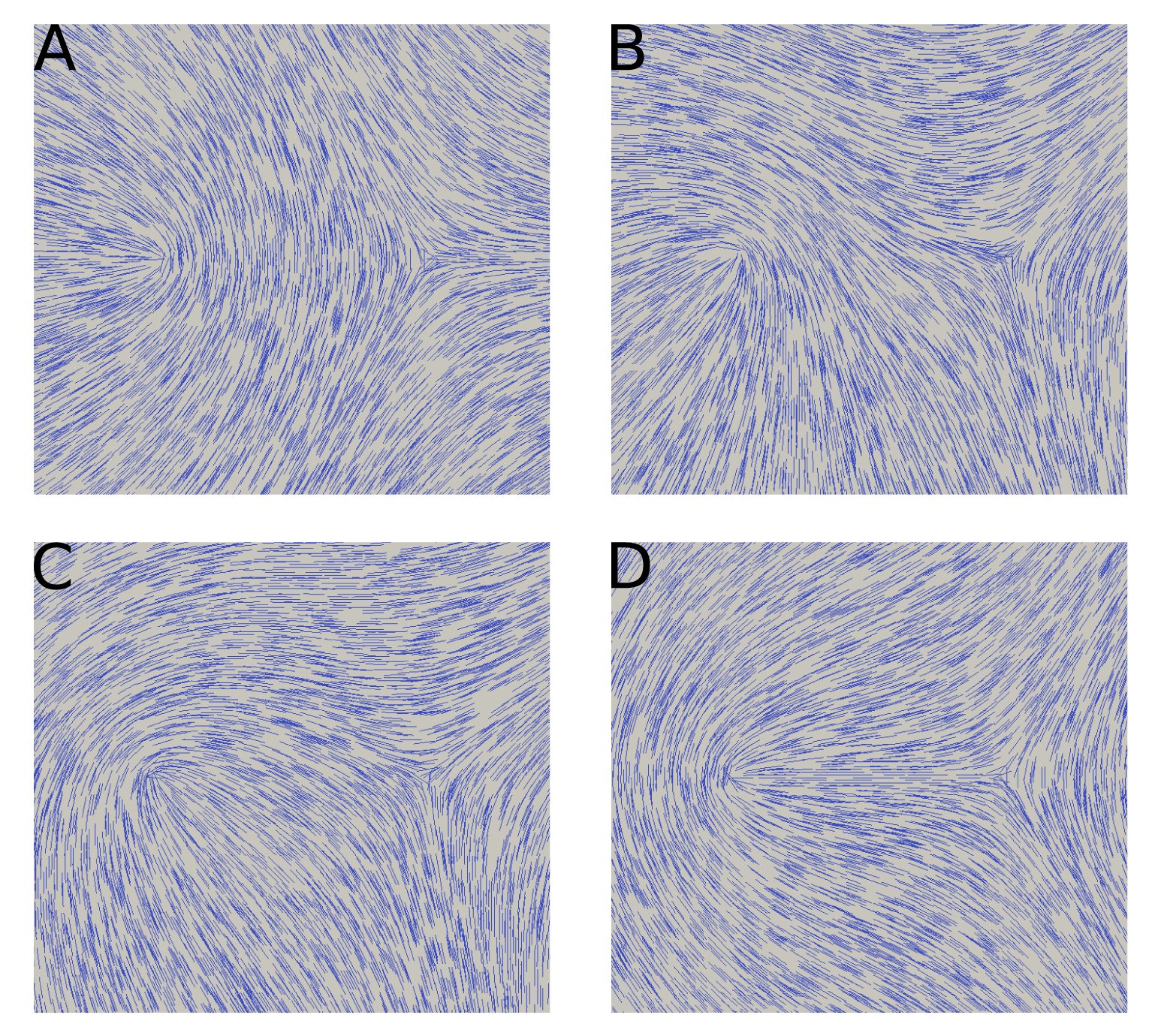}}
	\caption{\textbf{Director field associated with different defect orientations}. (A) $\Theta=0^\circ$; (B) $\Theta=45^\circ$; (C) $\Theta=135^\circ$; (D) $\Theta=180^\circ$.}
	\label{s6}
\end{figure}
\newpage


\subsection*{Supplementary Text}

In Fig. \ref{s5}, we show results from simulations that provide insights into how hydrodynamic effects influence defect morphology. The angle between the director and the flow direction for a strong shear flow is the Leslie angle $\psi$[33], as illustrated in Fig. \ref{s5}A. Such an angle and $\xi$ are related according to the following equation[46]:
$$
\xi \cos(2\psi)=3q_0/(2+q_0).
$$
In simulations, we tuned $\xi$ to control the Leslie angle $\psi$. As shown in Fig. \ref{fig2}D, the hydrodynamic flow associated with the active motion of the $+1/2$ defect gives rise to a strong shear flow. The directors respond to such a flow by aligning with it. We find that the lower $\psi$, the smaller $\kappa'$, consistent with the flow alignment picture.
To further examine the contribution of such hydrodynamic effects, we also perform a separate simulation in which the flow-alignment term ${\bf S}({\bf W},{\bf Q})$ in Eq.~7 is turned off. As shown in Fig. \ref{s5}B, the $\kappa'-\alpha$ curve becomes much flatter. This implies that flow-alignment effects are an important but not the only contributor to the change of defect morphology. The other contributor is the increase of activity-promoted bending in active systems.

We have performed simulations of defect dynamics with different orientations of the $+1/2$ defect, as illustrated in Fig. \ref{s6}. The orientation of the $-1/2$ defect in the initial state is chosen such that the overall elastic energy of the defect-pair system is a minimum. The dynamics of defect pair is determined by whether the defects annihilate in later stages of the simulation.

\end{document}